\newcommand{\secd}{\,{\rm s}}
\newcommand{\km}{\,{\rm km}}
\newcommand{\yr}{\,{\rm yr}}
\newcommand{\kms}{\km\secd^{-1}}
\newcommand{\masyr}{\,{\rm mas}\yr^{-1}}

\documentstyle[11pt,aaspp4,flushrt]{article}

\lefthead{Monnier et al.}
\righthead{VY CMa Outflow}

\begin{document}
\title{The last gasps of VY~CMa: Aperture synthesis and adaptive optics imagery}
\author{J. D. Monnier\altaffilmark{1}, P. G. Tuthill\altaffilmark{1},
B. Lopez\altaffilmark{2}, P. Cruzalebes\altaffilmark{3},
W. C. Danchi\altaffilmark{1}, and C. A. Haniff\altaffilmark{4} 
} 
\altaffiltext{1}{Space Sciences Laboratory, University of California, Berkeley,
Berkeley,  CA  94720-7450
}
\altaffiltext{2}{Observatoire de la C\^ote d'Azur, Departement Fresnel UMR 6528,
BP 4229, F-06034 Nice Cedex 4, France}
\altaffiltext{3}{Observatoire de la C\^ote d'Azur, Departement Fresnel UMR 6528,
Av. Copernic, F-06130 Grasse, France}
\altaffiltext{4}{Mullard Radio Astronomy Observatory, Cavendish Laboratory,
Madingley Road, Cambridge, CB3 0HE, UK}

%
%
%
%
\begin{abstract}
We present new observations of the red supergiant VY~CMa at
1.25\,$\micron$, 1.65\,$\micron$, 2.26\,$\micron$, 3.08\,$\micron$ and
4.8\,$\micron$.  Two complementary observational techniques were
utilized: non-redundant aperture masking on the 10-m Keck-I telescope
yielding images of the innermost regions at unprecedented resolution,
and adaptive optics imaging on the ESO 3.6-m telescope at La Silla
attaining extremely high ($\sim$10$^5$) peak-to-noise dynamic range
over a wide field.  For the first time the inner dust shell has been
resolved in the near-infrared to reveal a one-sided extension of
circumstellar emission within \mbox{0 \farcs 1 ($\sim$15\,R$_\star$)}
of the star.  The line-of-sight optical depths of the circumstellar
dust shell at 1.65\,$\micron$, 2.26\,$\micron$, and 3.08\,$\micron$
have been estimated to be 1.86$\pm$0.42, 0.85$\pm$0.20, and 0.44$\pm$0.11.
These new results allow the bolometric luminosity of VY~CMa to be estimated
independent of the dust shell geometry, yielding L$_\star \approx$ 2$\times
10^5\, $L$_\odot$.  A variety of dust condensations, including a large
scattering plume and a bow-shaped dust feature, were observed in the faint,
extended nebula up to 4\arcsec\ from the central source. While the origin of
the nebulous plume remains uncertain, a geometrical model is developed
assuming the plume is produced by radially-driven dust grains forming at a
rotating flow insertion point with a rotational period between
1200-4200\,years, which is perhaps the stellar rotational period or the
orbital period of an unseen companion.  
\end{abstract}


\section{Introduction}

VY~CMa (spectral type M5eIbp) is a very unusual star, displaying intense
radio maser emission from a variety of molecules, strong dust emission in the
mid-infrared, high polarization in the near-infrared, and large amplitude
variability in the visible.  Lada \& Reid (1978) estimated a distance of
$\sim$1.5~kpc for VY~CMa by establishing its association with a known
molecular cloud complex.  Strong independent confirmation of this distance
estimate has recently come from measuring the proper motions of long-lived
H$_2$O maser features in its outflow (\cite{marvel96}; \cite{ryc98}).  This
distance implies that VY~CMa is also one of the most intrinsically luminous
stars known in the galaxy, L$_\star \approx$ 5$\times 10^5\, $L$_\odot$
(\cite{sb96}).  Such high luminosity is only expected from massive stars
(M$_\star \approx 25$M$_\odot$) and the extremely low effective temperature,
T$_\star \approx  2800\,$K (\cite{sb96}), indicates that VY~CMa has $\sim$
10$^4$ years remaining before exploding as a supernova (\cite{bt82}).
Another sign of impending cataclysm is the extensive mass being lost by
VY~CMa into an optically thick circumstellar envelope.  The mass loss rate
for this star has been estimated using a variety of techniques (summarized in
[\cite{danchi94}]), yielding a consensus value of {$\dot{M}\approx$
2$\times10^{-4}\,$M$_\odot \,{\rm{yr}}^{-1}$}.

Previous observations of the circumstellar envelope at a variety of
wavelengths showed evidence of significant asymmetries.  The dense molecular
envelope surrounding the star supports strong maser emission of SiO, H$_2$O,
and OH, showing spatial and redshift distributions which conform to no simple
geometrical interpretation yet proposed (e.g. \cite{bm79}; \cite{marvel96};
\cite{ryc98}).  Optical observers have noted VY~CMa's peculiar one-sided
nebulosity for much of the century.  While these early observers described
VY~CMa as a binary or even multiple system (see [\cite{worley72}] for some
discussion), Herbig (1972), with the aid of polarizing filters, was able to
show definitively that the apparent companions were simply light scattered from
dust condensations in the nebula. Indeed, recent visible images from the
Hubble Space Telescope (\cite{kw98}) show a one-sided reflection nebula, with
a complicated arrangement of scattering features.  High values of
near-infrared linear polarization further suggest asymmetries in the
circumstellar emission at high spatial resolution, although no such
asymmetries had been directly detected until this work.  McCarthy (1979)
reported asymmetry in the mid-infrared emission of the dust shell,
interpreting this as evidence for thermal emission from a disk-like
structure.  Another important observational fact is that VY~CMa is an
irregular variable star showing 1-3~mag optical variation on the time scale
of $\sim$2000~days (\cite{marvel96}).  Recent changes in the near-infrared
polarization direction (\cite{maihara76}) and mid-infrared spectrum
(\cite{mgd98}) reveal that VY~CMa is evolving on a variety of time scales.

We present new infrared observations of VY~CMa at 1.25\,$\micron$,
1.65\,$\micron$, 2.26\,$\micron$, 3.08\,$\micron$, and 4.8\,$\micron$.  These
data detail new aspects of the dust shell, both from scattered light and
direct thermal dust emission, at unprecedented spatial resolution and dynamic
range.  After describing the novel observing techniques employed to make
these images possible, we attempt to synthesize a coherent history of the
VY~CMa's recent mass-loss.

\section{Observations}

\subsection{Keck-I observations}

Aperture masking interferometry was performed by placing custom-made plates
in front of the Keck-I secondary mirror, reducing light from the 10-m primary
mirror to a predefined set of subapertures.  An example of a 21-hole pupil
arrangement and its corresponding interference pattern can be found in
Figure~\ref{fig:golay}, where each hole is roughly one atmospheric coherence
patch in size (35~cm at $\lambda=2.2\,\micron$).  The introduction of such a
mask creates a series of overlapping two-hole interference patterns projected
onto the detector array, allowing the Fourier amplitude and phase of each
baseline can be recovered without the addition of ``redundancy'' noise (e.g.
\cite{baldwin86}).  By summing the atmospherically degraded Fourier phases
around each closed triangle, we obtain the {\em closure phase}, an observable
which is independent of the corrupting phase sheet above the subapertures
(e.g.  \cite{jennison58}).  In the bright source limit, non-redundant
aperture masking delivers maximal signal-to-noise ratio of the closure
phases, which is critical for reconstructions of non-centrosymmetric images.
For fainter sources, a doubly-redundant mask geometry in the shape of an
annulus was used (\cite{hb92}), balancing the desire for stellar flux with
that for limited baseline redundancy.  Further engineering and performance
details may be found in Tuthill et al. (1998).

Raw visibilities were calibrated for the mean telescope-atmosphere transfer
function by utilizing nearly contemporaneous observations of point-source
stars.  The observing regimen devoted an equal amount of time to such
calibration stars, interleaved with the science targets under
nearly-identical conditions.  Even so, seeing and other changes make the
calibration process notoriously difficult and mismatchs between source and
calibrator data introduced the greatest uncertainty into the final results.

Having obtained a set of calibrated visibilities and closure phases,
there are a number of image reconstruction algorithms which
minimize spurious artifacts such as sidelobes arising from non-uniform
or noisy sampling of the Fourier plane.  One of the most popular of
these, known as the ``Maximum Entropy Method (MEM)'' (\cite{gs84};
\cite{sivia87}), has been used here.  In order to check the
reliability of the reconstructions, the MEM results have been compared
with those from the CLEAN reconstruction algorithm (\cite{hogbom74};
\cite{cw81}; \cite{pr84}), which is also widely used in radio
astronomy.  Image orientation calibration and tests of the entire data
analysis pipeline were performed by observing sets of known, close
binary stars on each observing night.

VY~CMa was observed in both January and December of 1997 at Keck-I using the
Near Infrared-Camera (NIRC), a cryogenically cooled 256$\times$256~pixel InSb
array (\cite{ms94}).  The magnified plate scale of 20.3~milliarcseconds/pixel
was sufficient to Nyquist sample fringes formed by the longest baselines at
2.2\,$\micron$, however still finer fringes formed at shorter wavelengths
were undersampled (\cite{matthews96}).  The dates, filter wavelengths, mask
geometries, and integration times are detailed in Table~1.  The unresolved
star $\sigma$~CMa (spectral type M0Iab) was used for all Keck-I calibration.
Masking techniques only require $\sim$100 speckle frames of the calibrator
and source in order to produce high SNR maps of bright sources, such as
VY~CMa.  Typical uncertainties in the closure phases obtained with
non-redundant masks were $\pm$2\arcdeg.

\subsection{ESO 3.6-m observations}

Infrared observations of VY~CMa were also obtained at the European Southern
Observatory (ESO) 3.6-m telescope at La Silla, Chile in December 1996 and
January 1997 using the SHARPII+ and COMIC cameras.  The SHARPII+ camera
(\cite{lacombe95}), which is sensitive from 1-2.5\,$\micron$, contains a
256$\times$256 pixel NICMOS~3 detector array and the high resolution
35~milliarcseconds/pixel plate scale was utilized.  The COMIC camera
(\cite{lacombe95}) has a wider wavelength coverage (1.0-5.0\,$\micron$) and
was used to follow-up the SHARPII+ results and enable 5\,$\micron$
observing.  However, this camera is equipped with only a 128$\times$128 pixel
format array, yielding a smaller field-of-view but maintaining the same high
resolution plate scale.  Table~1 contains details of the observations.

VY~CMa and point-source calibrator observations were interleaved to calibrate
for changing seeing conditions and to fully characterize the faint wings of
the telescope point spread function.  $\omega$~CMa (spectral type B2IV-Ve)
was used as a calibrator source for both 1.25\,$\micron$ and 4.8\,$\micron$
observations.  The adaptive optics system was poorly adjusted while using the
SHARPII+ camera, resulting in uncertainty regarding the highest resolution
image features.

The strength of the ESO system is the ability to obtain very high dynamic
range images with nearly diffraction-limited resolution.  The peak-to-noise
ratio for the raw J-band images is $\sim$10$^5$.  A simple MEM deconvolution
algorithm (\cite{vl95}) was used to suppress the scattering halo from the
bright central source, however it is important to note that the prominent
features discussed here were easily observed in the raw images and the
deconvolution process yielded only minor adjustments to the relative flux
levels.  In addition to the uncertainties in the adaptive optics closed-loop
performance,  the extraction of truly diffraction-limited images was hampered
by the different spectral types of our source and calibrator targets, making
broad band observations difficult to calibrate; however, the full-widths at
half-maximum of the point-source calibrator images were always within 10\% of
those expected for a diffraction-limited 3.6-m telescope.  

\section{Results}

\subsection{Keck-I results}

Figure~\ref{fig:keck} shows image
reconstructions at 1.65\,$\micron$,
2.26\,$\micron$, and 3.08\,$\micron$ from two epochs; observing details can be found
in Table~1.  Since VY~CMa itself is completely unresolved by the
Keck-I telescope, we chose to smooth each image so that the FWHM
of the central, dominant component is somewhat smaller than the
Rayleigh criterion, $\Theta_{\rm{FWHM}}=1.22 \frac{\lambda }{D}$ 
The
filled circle in each image frame represents the FWHM of the central
component as displayed, 30~mas for the 1.65\,$\micron$ and
2.26\,$\micron$ images and 40~mas for the 3.08\,$\micron$
observations.  In addition to eliminating spurious features in the
reconstructions as judged by internal consistency from multiple data
sets, this process allows comparison of image contours generated from
pupil masks with slightly different maximum resolutions.  Image
structures which appear at or below the lowest plotted contours are the
result of systematic errors in the image reconstruction process, which
is fundamentally limited by the calibration of the atmopheric transfer
function.

The most dramatic feature in the high-resolution maps of
Figure~\ref{fig:keck} is the one-sided emission in the
near-infrared.  With the exception of a weak extension at
$\lambda=3.08\,\micron$, no circumstellar flux above 1\% of the peak was
detected to the north and north-east of the central source.  Although the
dust emission morphology is very similar at the three wavelengths, there are
some departures at 1.65\,$\micron$ where the scattering efficiency of the
grains is higher.  Unfortunately, fluctuations in the atmospheric conditions
limit the dynamic range in our final maps to $\sim$100, preventing detection
of low surface brightness features in the extended nebula.

The extended southern circumstellar emission is consistent with published
near-infrared polarization measurements.  The 1-3\,$\micron$ flux is
polarized with a position angle between 60\arcdeg and 105\arcdeg, with a
trend toward higher angles for shorter wavelengths.  Polarization angles and
polarization strengths both appear to be slowly increasing in time with
recent measurements giving linear polarizations of 7\% at 1.65\,$\micron$ and
2\% at 2.2\,$\micron$ (\cite{forbes71}; \cite{maihara76}; \cite{takami92}).
Assuming the polarization arises from Rayleigh scattering, these observations
suggest an asymmetric near-IR brightness distribution extending toward PA
170\arcdeg (or PA -10\arcdeg).  This is in close agreement with the new high
resolution results presented in Figure~\ref{fig:keck}.

The 10-m baselines afforded by the Keck-I telescope allow us to
discriminate stellar from non-stellar (extended) flux within our
5\arcsec$\times$5\arcsec field of view.  Fits to the high resolution
visibility data indicate the central source diameter is smaller than
20~mas; this is consistent with its bolometric luminosity and
estimated effective temperature which imply a stellar diameter between
10 and 16~mas (e.g., \cite{danchi94} and this paper).  Extended
emission (i.e., emission not originating from the unresolved central
component) contributes 79.0$\pm$1.0\%, 60.5$\pm$1.0\%, and
73.0$\pm$1.0\% of the total flux at 3.08\,$\micron$, 2.26\,$\micron$,
and 1.65\,$\micron$ respectively, appearing constant between January
and December 1997 to within uncertainties.  Note how the fraction of
the total flux arising from the extended component goes through a
minimum near 2.2\,$\micron$.  This behavior is expected when the dust
emission mechanism changes from being mostly thermal (in the red) to
mostly scattering (in the blue).

The dynamic range of $\sim$100 in the maps allows us to investigate
the mass loss history of VY~CMa for the last few hundred years.  The
bright knot of emission at PA 155$\arcdeg$ and separation $\sim$65 mas
appears in all the Keck-I maps and is near the expected dust
condensation radius, probably a clump of dust recently formed.  This
separation is slightly larger than the 50~mas dust shell inner radius
from spherically symmetric models of the 11\,$\micron$ emission at
maximum light (\cite{danchi94}).  The observed clump-star separation
is a function of wavelength and is smaller for bluer wavelengths.  In
addition, the PA of the clump is larger for shorter wavelengths, which
may also help explain the rotation of the polarization angle with
wavelength discussed above.  This behavior may arise from the
differing wavelength dependencies for thermal emission and scattering
of dust in the inner envelope.  Because the clump-star separation is
close to the diffraction limit, higher resolution images will be
necessary to eliminate possible mapping ambiguities in the clump
location.  Lastly, we note in passing that the 2.26\,$\micron$ and
3.08\,$\micron$ maps seem to exhibit a counter-clockwise spiral
structure.

\subsection{ESO 3.6-m results}

\subsubsection{1.25\,$\micron$ imagery}
The ADONIS adaptive optics system and the infrared cameras on the
ESO 3.6-m telescope allowed the recovery of nearly-diffraction limited
(90\,mas) images of a 6\arcsec$\times$6\arcsec\ field around VY~CMa with a
peak-to-noise dynamic range of $\sim$10$^5$.  Dust structures in the outflow
can be easily seen in scattered light at 1.25\,$\micron$, revealing a record
of the mass loss history of the last thousand years.

Figure~\ref{fig:sharp} shows a wide field image of the circumstellar
emission around VY~CMa, and one can see a number of interesting
structures.  A bow-shaped scattering feature lies 4\farcs 2 to the
south, with a brightness of approximately 0.003\% of the peak flux.  In
addition, an arcing plume of emission extending west-northwest is
observed with a peak brightness of about 0.01\% of the peak flux,
terminating abruptly 4\farcs 3.  This is the first high-resolution 
image of the ``curved nebulous
tail toward 290\arcdeg" described by Herbig (1972).  A similar sized
field towards the east of VY~CMa was also observed, but revealed no
nebulosity above 0.01\% of the peak.

A bright knot of emission is also present $\sim$1\arcsec\ from the
star at PA 220\arcdeg.  The peak emission from this SW knot is 0.5\%
of the peak flux and is nearly 2 orders of magnitude brighter than
other circumstellar features.  This feature has been previously
detected at 2.2\,$\micron$ by Cruzalebes et al.  (1998) in September
1994 using the same telescope with the COME-ON+ adaptive optics
system.  Other weak features present include a ridge of emission extending
from the star towards the SW knot and a north-south partial shell of emission
approximately 200~mas to the west of the central source.

A few detector artifacts must also be noted.  The high brightness of this
source caused the detector rows and columns sharing the central source region
to show artificially high counts.  Faint residual vertical and horizontal
stripes can be seen extending from VY~CMa caused by this effect.  Additional
spurious structure in the maps can be traced to diffraction spikes and
scattering from various optical surfaces.  Unfortunately, this scattering was
not identical for the source and calibrator and has led to some low
level miscalibrations, such as some faint emission within an arcsecond of the
source, which is especially obvious to the east and north (less than 0.01\%
of peak).  There is probably a similar amount of faint residual flux to the west
and south in the map, but it is not evident because of the bright dust
features present.

Figure~\ref{fig:comic} shows VY~CMa observed by the COMIC camera
when the adaptive optics system was performing well.  As can be seen,
there is noticably less emission observed within an arcsecond
(especially to the east and north), while the other major features
have remained largely unchanged.  This image has a smaller field of
view and has been included to show more detail of the nebula's inner
region.  The central source is partially resolved and slightly
elongated north-south, however a detailed fit to the size was not
attempted due to uncertainties in the adaptive optics calibration at
short wavelengths.  As with the SHARPII+ image in
Figure~\ref{fig:sharp}, one can easily see the bright knot of emission
to the SW, and a partial shell of emission to the west.  A recent
y-band (550~nm) image of VY~CMa taken by HST (\cite{kw98}) also shows
evidence for a ridge of scattering material about 200~mas to the west,
however other features observed here (e.g. the bright SW knot) do not
appear.  The HST images have significantly lower peak-to-noise ratio
($\sim$500) which may explain their absence.

\subsubsection{4.8\,$\micron$ data}

The analysis of the 4.8\,$\micron$ COMIC data of VY~CMa was hampered by poor
background subtraction for the point source calibrator.  Because low surface
brightness features within $\sim$1\arcsec\ may be indicative of sub-optimal
adaptive optics performance as well as circumstellar dust emission, the point
source response of the entire telescope must be characterized before certain
identification of dim, diffuse outflow condensations.  Unfortunately the
calibrator chosen ($\omega$~CMa) was not bright enough to allow high quality
background subtraction.  Hence we feel that we can only report thermal
structures at 1\% of the peak with high confidence.  The only such feature to
report is a $\sim$1\% brightness enhancement at 4.8\,$\micron$ which appears
at the same location as the bright scattering knot seen at 1.25\,$\micron$,
$\sim$1$\arcsec$ to the southwest of the central source.

At 4.8\,$\micron$, the central source of VY~CMa was partially resolved when
compared with the point source calibrator observed before and after.  To
investigate the high-resolution morphology of the compact core of thermal
emission, the data were analyzed utilizing the techniques of speckle
interferometry.  Frames were windowed to remove the relatively high background
fluctuations, then the mean power spectra were accumulated and averaged
together over many exposures.  Simple models of the brightness distribution
were then fit directly to the calibrated power spectra (or visibility
data).

Drifts in the AO response precluded a meaningful measurement of thermal
emission asymmetries in the compact emission core around VY~CMa, but
circularly-symmetric fits to the visibility data do estimate the emission
region's size.  Uncertainty estimations were performed by analyzing various
subsets of the 200~calibrator (T$_{\rm{int}}=$1.0\secd/frame) and 700~source
frames (T$_{\rm{int}}=$0.01\secd/frame) and Figure~\ref{fig:mband} shows the
results for the full data set.  Visibility data at short baselines are poorly
calibrated due to seeing/AO changes and the different spectral types of the
source and calibrator, a common artifact in speckle observations.  Model fits
were performed on data between 0.9\,arcsec$^{-1}$ and 2.7\,arcsec$^{-1}$, a
range chosen to exclude low spatial frequency corruption due to source
geometry, poor thermal background subtraction, low SNR data near the
telescope cutoff frequency and spectral type miscalibration.  A single,
circularly symmetric Gaussian brightness distribution was fit (allowing the
visibility at the origin to be a free parameter) yielding a FWHM of
163$\pm$8~mas, which is marginally consistent with previously published
4.8\,$\micron$ results of 137$\pm$21~mas by Dyck et al. (1984) and 152~mas by
Bensammar et al. (1985).  Possible systematic effects due to the different
spectral types of the source and calibrator stars introduce another
$\sim$10\% uncertainty.

While the best fit model shows systematic deviations from the data, the
observations are too uncertain to justify the fitting of more complicated
models.  However, to emphasize the need for higher resolution and better
calibrated observations, we note that the best fit model for a Gaussian +
point source yields a radically different dust shell geometry than the
single Gaussian fit, with 36\% of the flux in a point source surrounded by a
larger Gaussian with FWHM 268~mas.  Long baseline data at 4.8\,$\micron$
which resolves out the dust shell and measures the contribution of the
unresolved component are necessary to constrain circumstellar models
at this wavelength.

\section{Discussion}

We proceed by dividing the wealth of structure observed in the circumstellar
environment of VY~CMa into size scales from the smallest (highest resolution)
to most extended, discussing these in turn in the following subsections.

\subsection{Emission asymmetry within 10~R$_\star$}

Based on classical spherical outflow models, stars such as VY~CMa might be
expected to be surrounded by limb-brightened shells when observed in the
near-IR where the optical depth is lower than in the visible.  The
high-resolution images from the Keck-I telescope
(Figure~\ref{fig:keck}), however, bear no resemblance at all
to this preconception, but instead show emission to be one-sided,
inhomogeneous, and highly asymmetric.  Such structure may be interpreted in
terms of one of three scenarios: less dust on the northern side of the
nebula, high dust optical depth to the north of the star, or forward-
scattering from the dust particles in a tilted equatorial disk.

If the dust shell is optically thin at near-IR wavelengths then the lack of
emission to the north implies {\em less} or {\em cooler} dust.  As northern
dust would be in equal proximity to the star as the hot southern material, it
seems unlikely that the dust could be cool.  Furthermore, a dramatically
rarified or evacuated region to the north also seems implausible when we
consider the shape of the reflection nebula at bluer wavelengths.  The almost
total absence of scattered light to the north and north-east in the
1.25\,$\micron$ images in Figures~\ref{fig:sharp} and \ref{fig:comic} and
also for images in the visible (\cite{herbig72}; \cite{kw98}) argues strongly
that thick obscuration blocks the observer's line of sight into the northern
region of the nebula.  H$_2$O masers present to the north and
east of the VY~CMa also indicates high dust densities in these locations
(\cite{ryc98}; \cite{marvel96}).

Alternatively we consider the case for an inner dust shell which is optically
thick even at our reddest wavelength of 3.08\,$\micron$.  In this case,
emission morphology traces lines-of-sight which can penetrate into hot, dusty
regions before encountering unity optical depth.  While spherically symmetric
outflows would appear as centrally-peaked shells, axisymmetric circumstellar
density distributions can look quite different depending on the viewing angle
(\cite{er90}; \cite{lml95}).  If the equatorial plane contains higher density
material than that found near the poles, then the one-sided emission observed
in the near-IR arises from warm dust in a polar region tilted somewhat
south-southwest from our line of sight, a viewing angle consistent with the
large scale scattering asymmetry in the visible (the one-sided nebulosity).
Furthermore, since the appearance of the object is determined to a great
degree by optical depth effects, it is not surprising that the overall
thermal emission extension occurs in roughly the same direction for
3.08\,$\micron$ as for the scattered light asymmetry observed at
1.65\,$\micron$.  The only 10\,$\micron$ observations of VY~CMa to detect
asymmetry were by McCarthy (1978) who found an extension in the NE/SW
direction. This is consistent with a disk-like dust distribution of high
optical depth with warm, evacuated polar regions, especially if large dust
grains are present enhancing mid-infrared scattering.

As a third scenario, the forward-scattering of stellar light by dust
contained in a tilted equatorial disk may contribute to the one-sided
appearance of the nebula.  If one assumes that the VY~CMa dust envelope is,
in a first approximation, a disk-like structure with its south-west side
tilted towards at us, then the high surface brightness of the SW side may be
related to enhanced small-angle, forward-scattering by large dust particles.
Such an effect has been observed in the circumstellar disk of UY~Aur
(\cite{close98}) where the ratio between the bright side and faint one is
about 10. Although the ratio between the bright and faint lobes is larger
than 100 for VY~CMa, the presence of larger dust particles and a smaller disk
inclination angle may help explain this difference.

The line-of-sight extinction due to the circumstellar shell and
intervening interstellar dust can be estimated by using a combination
of published infrared photometry, standard near-IR optical dust
constants, and the high-resolution interferometric observations
presented here.  The interferometric data and photometry allow us to
determine the absolute amount of flux at each wavelength which is
coming from the star, separating out the circumstellar emission.  By
using an established extinction law (\cite{ohm92}) and published
near-infrared photometry of VY~CMa near maximum light
(\cite{lebertre93}), the 2800~K blackbody spectrum of the star can be
reddened until it matches the observed flux ratios of the star at
1.65\,$\micron$, 2.26\,$\micron$, and 3.08\,$\micron$.  When using
three wavelengths, this procedure may not have a solution, which would
then imply the existence of errors in the flux measurements, the dust
constants, or the effective stellar temperature.  However, this
procedure does yield a unique solution when applied to VY~CMa,
indicating that the set of initial assumptions are indeed
self-consistent.  The line-of-sight optical depths (which includes the
dust shell as well as interstellar extinction) for 1.65\,$\micron$,
2.26\,$\micron$, and 3.08\,$\micron$ are 2.1, 1.0, and 0.53.  The
bolometric luminosity of VY CMa can also be estimated by dereddening
the observed stellar flux without additional assumptions regarding the
geometry of the dust shell.  Assuming a distance of 1500~pc, this
procedure yields L$_\star \approx$ 1.7$\times 10^5\, $L$_\odot$
(T$_\star \approx$ 2800\,K; R$_\star \approx$ 5.5~mas $\approx$
8.3~AU).  The uncertainties in these values are dominated by
systematic uncertainties relating to the choice of dust constants and
the effective temperature of the star.  Calculations done using other
dust constants (\cite{dl84}; \cite{dp90}) and with effective
temperatures as hot as 3500~K, indicate an uncertainty of $\sim$20\%
in the values of the line-of-sight optical depths, and that the
L$_\star$ may be as high as 3.0$\times 10^5\, $L$_\odot$.  This
determination of L$_\star$ is somewhat smaller (by factor of $\sim$2)
than the bolometric luminosity found for the spherical models by
Le Sidaner \& Le Bertre (1996).  However spherical
models of axisymmetric dust shells overestimate the bolometric
luminosity when the viewing angle looks into less dense polar regions
rather than through the optically thicker equatorial plane, and this
may explain the discrepancy.

We must correct for interstellar extinction in order to estimate
the optical depth of the circumstellar shell.  Efstathiou \& Rowan-Robinson
(1990) calculated various circumstellar dust shell models of VY~CMa using
interstellar extinction, A$_V$, from between 1.2 and 1.5.  Despite the
uncertainty in these estimates, we adopt $A_V=1.35$ and use the Mathis (1990)
extinction law to correct the optical depth determinations for interstellar
extinction.  The line-of-sight optical depths of the circumstellar dust shell
around VY~CMa at 1.65\,$\micron$, 2.26\,$\micron$, and 3.08\,$\micron$ then
become 1.86$\pm$0.42, 0.85$\pm$0.20, and 0.44$\pm$0.11.  These values
are consistent with the hypothesis that the optical depth in the equatorial
plane is roughly equal to or greather than unity at 3.08\,$\micron$, as long
as the dust density is enhanced in an equatorial plane inclined to our
line-of-sight.

High-resolution maps at our three infrared colors show similar structure,
however marked differences exist, particularly between the 1.65\,$\micron$
maps of Figures~2a \& 2b and those of Figures~2c-f
further in the red.  Such wavelength dependent changes are
the result of a number of physical phenomena.  Regions of dust will yield
varying spectral contributions depending on the local temperature, the
importance of scattering, and the overall optical depth; all of these effects
are strongly wavelength dependent.  
Although we could make some attempt to
disentangle these three effects with our multi-wavelength maps, such an
effort is probably more profitable in the context of the construction of a
detailed, self-consistent radiative transfer model of this star; this is left
for future work.

While no relative motion of the dusty clumps was detected between the
January and December of 1997, follow-up high-resolution imaging of
this source over the coming decades may show motion of individual
clumps of emission, if they represent dust clouds being accelerated
and driven away from the star.  If the dusty clumps have an outflow
velocity typical of the OH and H$_2$O molecules ($\sim 33 \kms$), this
4.6$\masyr$ outward motion could be detected with a temporal baseline
of only a few years.  

\subsection{Southwestern nebulosity and bright knot}

Turning our attention to the more extended structure revealed by adaptive
optics, circumstellar emission at 1.25\,$\micron$, 2.2\,$\micron$ and
4.8\,$\micron$ within the inner arcsecond (or $\sim$150\,R$_\star$) of VY~CMa
is dominated by the southwestern nebulosity and bright knot (see
Figure~\ref{fig:comic} and Cruzalebes et al. [1998]).  The direction of this
extension is roughly perpendicular to the possible equatorial plane as
deduced from the one-sided reflection nebula seen in the visible.  This
near-IR, southwestern nebulosity supports the interpretation that light is
preferentially escaping through less dense polar regions of a roughly
axisymmetric dust distribution inclined towards the southwest.  However we
note the position angle of the knot is $\sim$30\arcdeg\ different from the
extension observed at high resolution in the inner dust shell at
1.65\,$\micron$ (see Figures~2a \& 2b).  An overall axisymmetric dust
distribution with an axis lying roughly NE/SW would allow observations of
dust emission only in the forward-facing, SW polar lobe.  Hence, dust
inhomogeneities observed at higher resolution (see
Figure~\ref{fig:keck}) may reflect the
chaotic nature of local dust formation and may not be directly associated
with the global symmetry of the dust shell.

The bright knot at 1\arcsec\ to the southwest may be partly due to an
excess of stellar illumination through a hole in the inner dust shell.
The relatively narrow line of emission connecting the knot with the
central region lends some support for this conjecture, however this
may alternatively indicate a bridge of material extending from the
star towards the bright knot.  Figure~\ref{fig:comic} also summarizes
a record of ``binary companion'' observations in the visible
stretching back over 70\,years (\cite{wal78}).  The current location
of the SW knot can be seen to lie along an extrapolated locus of these
earlier points.  Wallerstein (1978) attempted to explain this motion
with a ``rotating hole'' hypothesis wherein the orbital motion of hole
in the inner dust shell results in a moving ``searchlight'' beam.
Rotation rates derived from this model are too high to be reconciled
with single-star evolution of a red supergiant, implying the presence
of a close binary companion or that VY~CMa is actually a pre-main
sequence object still in possession of significant disk angular
momentum.

Firm conclusions on the physical meanings of these features are
difficult to draw mainly due to the many uncertainties of the dust
shell geometry.  For instance, the three-dimensional location of the
SW knot is very uncertain, hence the knot could represent scattering
from a dust cloud existing above the pole or not far off the
equatorial plane.

\subsection{Faint dust structures in the outer envelope}

Dust features were detected in scattered light at 1.25\,$\micron$ at
distances up to 4\arcsec\ from the central source, which corresponds
to a separation of $\sim$600\,R$_\star$.  The two main features observed at
this size scale are the southern bow-shaped feature and the ``curved,
nebulous tail'' stretching to the northwest in Figure~\ref{fig:sharp}.

The narrow arc or bow-shaped feature to the south may have originated
from a partial shell of dust formed during an asymmetric ejection
event, or alternatively as a cloud of higher velocity dust which was
subsequently distorted by interactions with preexisting circumstellar
material.  This dust material may be embedded in the expanding
envelope around this star; if so, the dust cloud's velocity can
be estimated from the expansion velocity of molecular gas (OH, H$_2$O)
in this region, $33-38 \kms$ (\cite{bjs83}; \cite{ryc98}).  This
would date the time of ejection as $\sim$1000 years ago.

In order to understand the origin of the dramatic western arc seen at
1.25\,$\micron$, we must consider two classes of explanations.  The
first class would not involve the actual transport (or collimated
outflow) of dust in forming the arcing shape.  This class of theories
would hold that the circumstellar envelope has been severely
disrupted, perhaps from vestigial pre-main-sequence material or the
passage of a wide binary through the envelope.  The nagging
possibility that VY~CMa is actually a pre-main-sequence object falls
into this catagory (\cite{lr78}; {\cite{kw98}). Local obscuration and
unusual illumination scenarios play a role here in shaping the
appearance of the circumstellar environment.  Until additional
observations offer further support for such scenarios, we decline
further speculation along these lines.

The second class of explanations involves the actual transport of dust
in a collimated outflow.  This idea is supported by the large contrast
of the plume emission with respect to the surrounding nebulosity.  The
1-2 orders of magnitude difference suggest that the density or albedo
of the material along the plume is significantly higher.  Assuming the
plume material did indeed originate at the star and is embedded in the
outflow, we further consider two cases depending whether the flow is
``fast'' or ``slow.''  If the plume arises from a ballistic jet of
high-velocity material ejected from the star, one would expect shock
waves to form in the outflow.  In order to investigate the ballistic
jet hypothesis for the plume a coronagraphic study was undertaken at
Keck-I in December 1997 to search for line emission.  No emission
above the continuum was observed in filters sensitive to the H$_2$
($\nu$=1-0) and H$_2$ ($\nu$=2-1) lines, transitions indicative of
shocked hydrogen.  The coronagraphic study did confirm the blue
spectrum of the plume (as expected from scattering from grains) by
mapping the [1.6\,$\micron$]/[1.2\,$\micron$] color.

On the other hand, if the dust in the plume has the ``slow'' outflow velocity
typical of the OH and H$_2$O the molecules, this
4.6$\masyr$ radial motion implies that the plume process ``turned on''
roughly 1000~years ago.  The long extension of the plume suggests that
the ejection direction maintained itself, except for some apparent
rotation, during most of the time since the beginning of the flow.
We pursue this possiblity below.

\subsection{Geometrical model of plume}
A simple geometrical model has been constructed to test the hypothesis
that the plume represents radially-driven dust grains produced at a
rotating, or precessing, flow insertion point; for simplicity, we
assume the plume lies in a plane.  The velocity of the dust as a
function of distance from the star has been based on the proper motion
study of the H$_2$O masers by Richards, Yates, \& Cohen (1998).  The
dust is assumed to have an 8.5~km/s outflow velocity at 75~mas,
increasing linearly to 33.0~km/s by 400~mas and staying constant
thereafter, although our results are not critically dependent on the
velocity law close to the star.  We can now fit models based on three
input parameters: the initial ejection direction, the period of
rotation, and the inclination of the rotation axis (i.e., the
viewing angle).

The model outflow was constrained to fit the outer ridge of the plume
between 2\farcs 5 and 4\farcs 1 and to pass through the SW knot
located approximately 1\arcsec\ away from the star.  The initial
ejection direction was fit to the direction of the near-infrared
curved extension seen towards the south in the Keck-I
images (Figures~2e \& 2f).  The best fit plume model appears in
Figure~\ref{fig:sharp}, which contains the shape of the outflowing
plume structure as well as the projection of the rotation equator and
the polar axis.  The model fit indicates a rotational period between
1200-4200~years, with an equatorial plane inclined
51\arcdeg$\pm$5\arcdeg\ out of the plane of the sky and the North (or
South) polar axis projected onto PA 200\arcdeg$\pm$15\arcdeg\, with a
180\arcdeg\ ambiguity (i.e., the fit can not tell us which lobe is
facing towards the observer).  With this geometry, the estimated time
that the flow began is 950~years ago, and the flow insertion
point is rotating clockwise when looked upon from the north pole.
Unfortunately, the largest uncertainties in this fit are in the model
assumptions themselves.  The identification of the spiral-type
morphology passing inwards from the plume through the SW knot and on
in the Keck-I images is highly speculative, hence our model parameters
should be considered tentative.

In spite of such uncertainties, our simple empirical model shows a number of
promising properties, and furthermore serves as a useful starting point in
exploring the details of the circumstellar morphology.  The model is
consistent with the assumption that the axisymmetry of the one-sided
reflection nebula is due to an equatorial density enhancement with the
equatorial plane inclined to the line-of-sight in such a way that scattering
to the north and northeast is not seen.  Efstathiou \& Rowan-Robinson (1990)
constructed a radiative transfer model of VY~CMa based on a flared disk
geometry with an equatorial plane inclination angle of $\sim$47\arcdeg .
This is entirely consistent with the plume model developed above, although
their model results depend greatly on the assumed density distribution of the
disk.  In addition, the VY~CMa's derived rotational period of $\sim$2700~yr
is the right order of magnitude for evolution from an O-star progenitor
(\cite{heger98}).

However, there is no obvious explanation for how the outflow could
persist for 1000~years. For instance, the evolutionary time scale for
the convective elements within the photosphere has been found to be
less than one year (\cite{thb97}; \cite{wdh97}).  However this does
not preclude the existence of long lasting surface inhomogeneities
supported by magnetic fields, rotation, or some other complex
phenomenon (e.g. Jupiter's Red Spot).  If such a localized hot-spot or
cool-spot exists, it may catalyze excess mass-loss or the creation of
dust with a peculiar grain size distribution or chemical composition.
The flow insertion point for this dust would be seen to turn at
approximately the stellar rotation rate.

Alternatively, VY~CMa may have experienced a dramatic stellar disruption
about 1000~years ago which resulted in significant asymmetrical mass loss.
This would link the origin of the southern bow-shaped dust feature with the
beginning of the plume flow, since they are at the same distance from the
central source.  Such a disruption could occur if the outer surface layers
become unstable to large amplitude pulsations (\cite{heger97}) or if the star
undergoes an asymmetrical shell flash, although such internal, nova-like
outbursts are not expected for stars more massive than $\sim$11\,M$_\odot$
(\cite{iben98}).  This ejected material may persist in the inner envelope
for years until the gas either disperses, falls back to re-join the
photosphere, or nucleates into dust grains and forms the arc.  
One problem with this picture is that a large quantity of gas would be
required to form the high contrast plume seen in the scattered light, the
plume being at least 10-100~times brighter than its surrounding nebulosity.
Such increased gas density in the upper atmosphere would cause a large
pressure imbalance, spreading the gas around the star within
$\sim$100~years (local sound speed is $\sim 3\kms$).
However, the large scattering contrast in the plume could also be
caused by the presence of high albedo dust grains instead of 
higher dust density.  The scattering efficiency
of grains can be dramatically increased by changing their size or their
chemical composition (\cite{pp83}).  

One last interpretation for the 2700~yr period would be that it
represents an orbital time period for a close binary companion.
Between 31\% and 62\% of O-stars (\cite{garmany80}; \cite{stone81}),
progenitors of red supergiants, are found to be in binary systems.
A low-mass companion located approximately 400~mas from 
VY CMa would have an orbital period of about 2700~years, but it is is
unknown whether such a wind/binary interaction could produce the high
contrast plume observed.

\section{Conclusions}

We have reported new high angular resolution and high dynamic range
observations of the circumstellar outflow of VY~CMa.  In addition to
confirming previously observed features, several new structures were
discovered.  Diffraction-limited observations with the Keck-I telescope allow
the star and inner dust shell to be observed through the dusty envelope,
revealing one-sided emission within 10~stellar radii.  Combining these
results with published photometry and existing optical dust constants, the
line-of-sight optical depths of the circumstellar dust shell at
1.65\,$\micron$, 2.26\,$\micron$, and 3.08\,$\micron$ were estimated to be
1.86$\pm$0.42, 0.85$\pm$0.20, and 0.44$\pm$0.11.  In addition, these new
results allow the bolometric luminosity of VY~CMa to be estimated independent
of dust shell geometry, yielding L$_\star \approx$ 2$\times 10^5\,
$L$_\odot$.

The adaptive optics system on the 3.6-m telescope 
has mapped the circumstellar
envelope out to $\sim$4\arcsec\ from the faint stellar flux scattered
off grains in the outflow.  This combination of high spatial
resolution ($\sim$90 mas) and high dynamic range ($\sim$10$^5$) was
critical in resolving a high contrast, scattering plume and a narrow,
bow-shaped feature in the extended circumstellar environment.

The simplest density distribution consistent with the present data is an
axisymmetric one with an equatorial density enhancement, the pole oriented
roughly NE/SW with the SW lobe tilted towards us.  This geometry attributes 
the lack of observed emission to the north and northeast to
obscuration by dust near the equatorial plane.  The clumpy near-IR emission
on spatial scales of a few stellar diameters indicates the local mass-loss
processes are strongly inhomogeneous.  Although the present site of dust
formation within 10~R$_\star$ (PA $\sim$155\arcdeg) does not strictly conform
to the overall geometry of the dust shell, time-averaged axisymmetry around a
NE/SW axis is still possible and is most consistent with the full set of
observations.

The curved, nebulous morphology of the dust plume extending to the northwest
is consistent with radially-driven dust grains formed at a rotating flow
insertion point due to asymmetric mass-loss or a binary
companion.  A simple geometrical model indicates a rotational or orbital
period of $\sim$2700~years.  The outer edge of the plume and the southern,
bow-shaped feature lie approximately the same distance from the star
suggesting a common origin in a stellar disruption $\sim$950~years ago.  If
such a disruption could liberate carbon-rich material from near the core, the
long-life of the plume and its high-contrast with respect to the surrounding
nebulosity might be explained as due to albedo variations of chemical
origin.

The inner envelopes of late-type stars are proving much more complicated and
exciting than expected.  The clumpy scattering features in
high-resolution HST and Keck-I images, the nebulous curving plume and the
southern, bow-shaped feature embedded in the outflow, non-radial maser proper
motions (\cite{ryc98}; \cite{marvel96}), time evolution of the polarization
direction (\cite{maihara76}), and radical changes in mid-infrared spectrum
(\cite{mgd98}) all testify to the presence of
chaotic and violent dust formation processes around VY~CMa which occur on
time scales of decades to millenia.  Effects of stellar hotspots
(\cite{thb97}), high magnetic fields (\cite{mpp94}; \cite{kd97}), rotation
(\cite{hl98}), large amplitude pulsations (\cite{heger97}), and close binary
companions (\cite{morris80}) may all play important roles in producing the
thick circumstellar envelopes observed during the dying days of red
supergiants and subsequently illuminated in the afterglow of supernova
explosions.

\acknowledgments

{We would like to thank Devinder Sivia for the maximum-entropy mapping
program ``VLBMEM,'' which we have used to reconstruct
diffraction-limited images.  We wish to acknowledge the Keck
Observatory personnel, especially Peter Gillingham,
who have supported the aperture masking
experiment and to whom we owe great thanks.
JDM would also like to thank Chris Matzner for helpful discussions.
This research has made extensive use of the SIMBAD
database, operated at CDS, Strasbourg, France, and NASA's Astrophysics
Data System Abstract Service.  Some of the data presented herein were
obtained at the W.M. Keck Observatory, which is operated as a
scientific partnership among the California Institute of Technology,
the University of California and the National Aeronautics and Space
Administration.  The Observatory was made possible by the generous
financial support of the W.M. Keck Foundation.  The results presented
here have been partly based on observations collected at the European
Southern Observatory, La Silla (Chile).  CAH is grateful to the Royal
Society for continued financial support.  This work is a part of a
long-standing interferometry program at U.C. Berkeley, supported by
the National Science Foundation (Grant AST-9315485 and AST-9731625) 
and by the Office of Naval Research (OCNR N00014-89-J-1583).  In addition,
scientific exchanges with the Observatoire de la C\^ote d'Azur were supported
in part by a 1997 France-Berkeley Fund grant to WCD.  } 
\pagebreak 

\clearpage

\begin{deluxetable}{lcccclll}
\small
\tablewidth{0pt}
\tablecaption{Journal of Observations}
\tablehead{
\colhead{Date}  &   \colhead{$\lambda$}  & 
\colhead{$\Delta\lambda$} &  \colhead{T$_{\rm{int}}$}& \colhead{Number of} & \colhead{Telescope} & \colhead{Camera} &
\colhead{Comments}  \\ 
\colhead{(U.T.)} & ($\micron$) & ($\micron$) & (ms) & Frames & & &  }
\startdata
1996 December 29 & 1.253 & 0.296 & 100 & 480 & ESO 3.6-m & SHARPII+ & Adaptive optics sub-optimal \\
1997 January 2   & 1.259 & 0.229 & 2000 & 200 & ESO 3.6-m & COMIC    & With ADONIS adaptive optics \\
1997 January 2   & 4.832 & 0.590 & 10 & 700 & ESO 3.6-m & COMIC    & With ADONIS adaptive optics \\
\\
1997 January 30  & 1.647 & 0.018 & 134 & 100 & Keck-I    & NIRC     & With 15-hole aperture mask \\
1997 January 30  & 2.260 & 0.053 & 134 & 100 & Keck-I    & NIRC     & With 15-hole aperture mask \\
1997 January 30  & 3.083 & 0.101 & 134 & 100 & Keck-I    & NIRC     & With 15-hole aperture mask \\
\\
1997 December 17 & 1.647 & 0.018 & 149 & 100 & Keck-I    & NIRC     & With annulus aperture mask \\
1997 December 19 & 1.647 & 0.018 & 149 & 100 & Keck-I    & NIRC     & With annulus aperture mask \\
1997 December 19 & 2.260 & 0.053 & 149 & 100 & Keck-I    & NIRC     & With 21-hole aperture mask \\
1997 December 19 & 3.083 & 0.101 & 149 & 100 & Keck-I    & NIRC     & With 21-hole aperture mask \\
\enddata
\end{deluxetable}

\clearpage      
\begin{figure}
\figurenum{1}
\plotfiddle{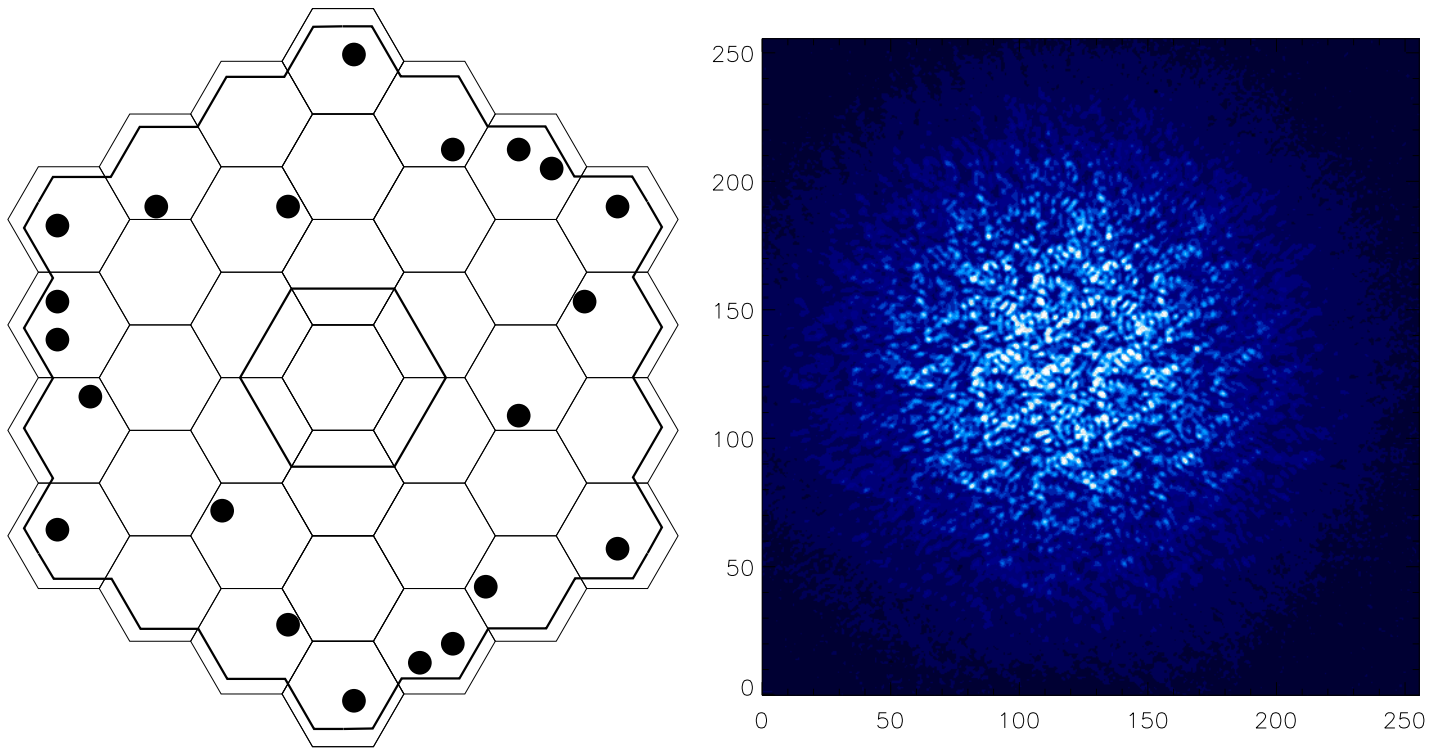}{4.0 in}{-90.0}{75.0}{75.0}{-300}{400}
\caption{\label{fig:golay}
({\em left panel}) The pupil arrangement of a 21-hole, non-redundant
aperture mask projected onto the hexagonal primary mirror of Keck-I.
({\em right panel}) A single speckle frame observed while utilizing the
aperture mask is shown to the left. The wavelength of observation was
3.08\,$\micron$ and the integration time was 0.149\,s.  The horizontal
and vertical axes are labeled in units of pixels, the plate scale being
20.3\,milliarcseconds per pixel.}
\end{figure}

\begin{figure}
\figurenum{2}
\plotfiddle{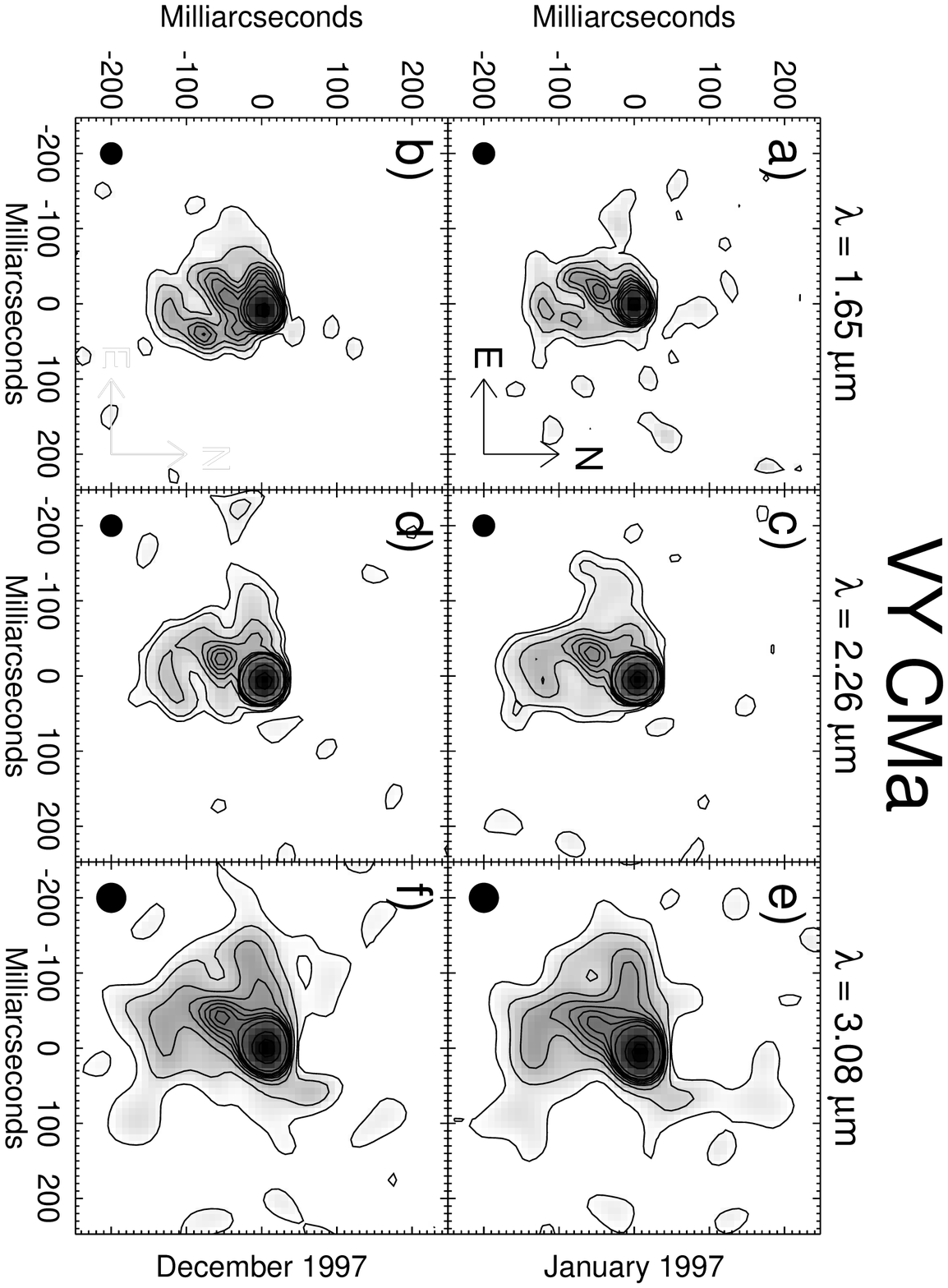}{4.0 in}{90.0}{75.0}{75.0}{300}{-40}
\caption{\label{fig:keck}
{\em Panels a \& b:} 
Image reconstructions of VY~CMa observed at 1.65\,$\micron$
in January and December of
1997.  The unresolved central source is displayed with a FWHM of 30~mas,
as indicated by the ``beam'' circle in the lower-left corner of each panel.
The contour levels are 1.5\%, 3.0\%, 5.0\%, 7.0\%, 9.0\%, 11.0\%, 15.0\%, 
20.0\%, 30.0\%, and 80.0\% of the peak. 
{\em Panels c \& d:}
Image reconstructions of VY~CMa observed at 2.26\,$\micron$
in January and December of
1997.  The unresolved central source is displayed with a FWHM of 30~mas,
while the contour levels are 0.7\%, 1.0\%, 2.0\%, 3.0\%, 4.0\%, 5.0\%,
6.0\%, 10.0\%, 30.0\%, and 70.0\% of the peak.
{\em Panels e \& f:}
Image reconstructions of VY~CMa observed at 3.08\,$\micron$
in January and December of
1997.  The unresolved central source is displayed with a FWHM of 40~mas,
while the contour levels are 1.0\%, 2.0\%, 4.0\%, 6.0\%, 8.0\%, 10.0\%, 
12.0\%, 14.0\%, 20.0\%, 50.0\%, and 80.0\% of the peak.
}
\end{figure}

\begin{figure}
\figurenum{3}
\plotfiddle{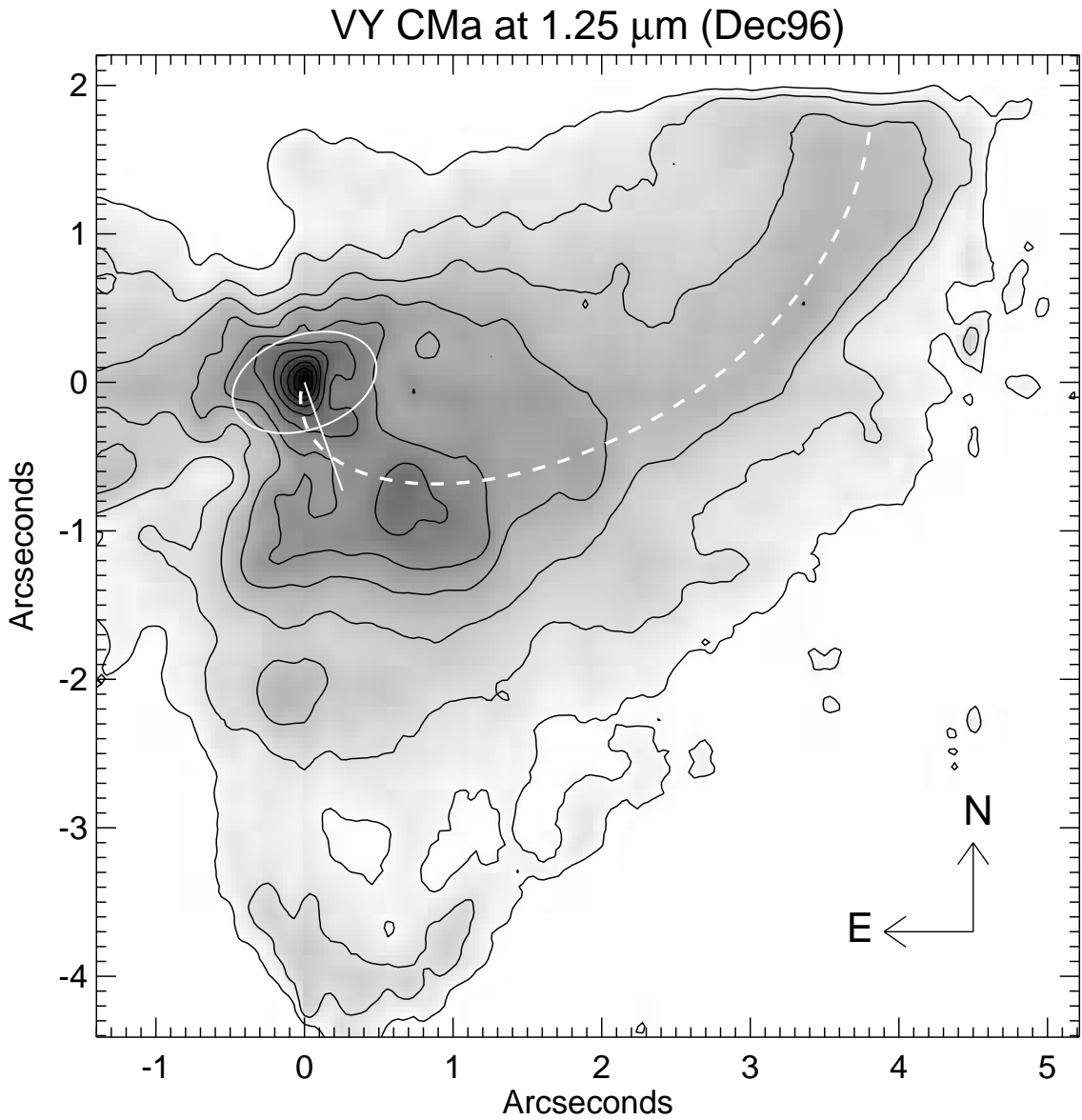}{4.0 in}{90.0}{90.0}{90.0}{400}{-70}
\caption{\label{fig:sharp}
Adaptive optics image of VY~CMa observed at 1.25\,$\micron$ in
December 1996 using the SHARPII+ camera.  The overplotted solid and dashed white lines
indicate the rotation geometry and outflow trajectory
for the best-fit plume model discussed in \S 4.4.
The contour levels are 0.001\%, 0.003\%, 0.010\%, 0.032\%, 0.10\%, 
0.316\%,  1.0\%, 3.16\%, 10.0\%, and 31.6\% of the peak.
}
\end{figure}

\begin{figure}
\figurenum{4}
\plotfiddle{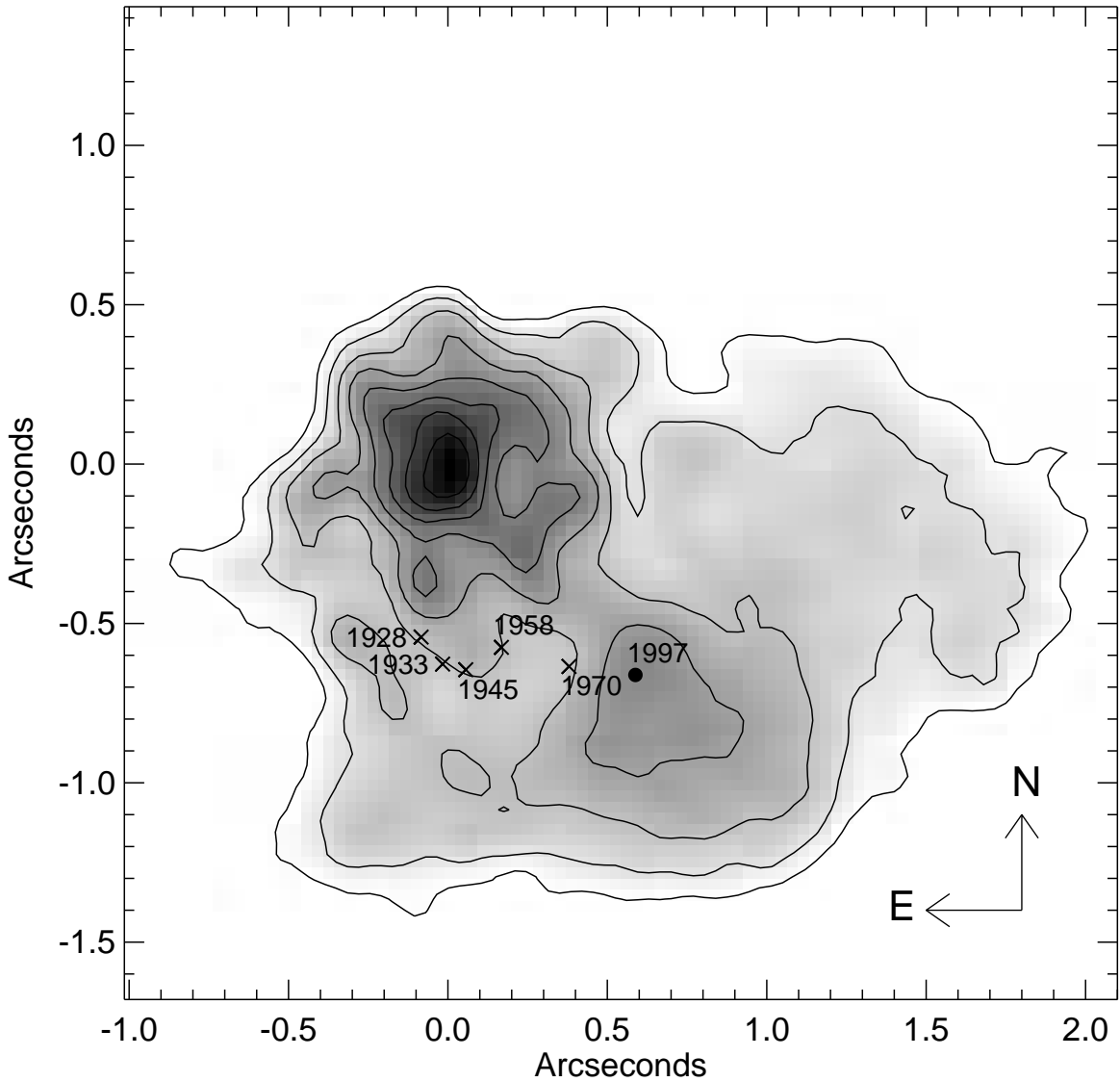}{4.5 in}{90.0}{90.0}{90.0}{400}{-70}
\caption{\label{fig:comic}
Adaptive optics image of VY~CMa observed at 1.25\,$\micron$ in January
1997 using the COMIC camera.  The cross symbols and accompanying dates
represent the mean location of VY~CMa's ``companion'' as observed in
the visible by the U.~S. Naval Observatory over the last 70 years (see
[\cite{wal78}] for a summary of these observations).  The predicted position of
the ``companion'' in 1997 has been extrapolated from the historical data
and included as a filled circle.  The contour levels are 0.010\%,
0.032\%, 0.10\%, 0.316\%,  1.0\%, 3.16\%, 10.0\%, and 31.6\% of the
peak.}
\end{figure}

\begin{figure}
\figurenum{5}
\plotfiddle{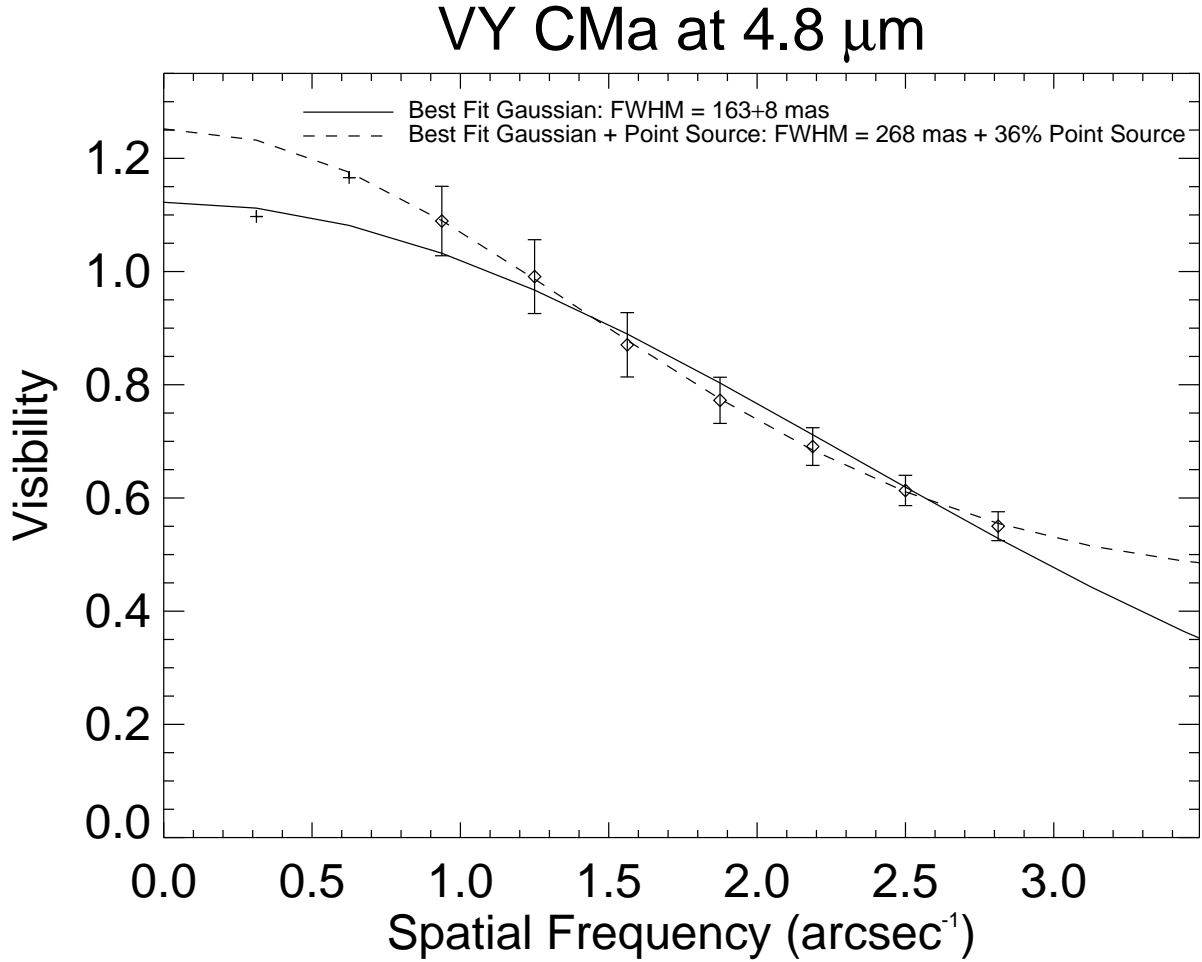}{4.0 in}{90.0}{75.0}{75.0}{280}{-40}
\caption{\label{fig:mband}
Azimuthally-averaged visibility measurements of VY~CMa 
at 4.8\,$\micron$ observed with the COMIC camera in January 1997. 
The best fit circularly-symmetric Gaussian distribution, FWHM 163$\pm$8~mas,
is included on the figure as a solid line.  The plus symbols are
data points not used for the fitting.
To illustrate the importance of acquiring longer baseline information,
an alternative model consisting of a Gaussian distribution
with FWHM 268~mas and a point source is shown as a dashed line.
}
\end{figure}
\end{document}